\documentclass[12pt,preprint2]{aastex}

 \usepackage{epstopdf}
\usepackage{url}
\usepackage{graphicx}
\usepackage{longtable} 

\usepackage{array}
\newcolumntype{H}{>{\setbox0=\hbox\bgroup}c<{\egroup}@{}}

\shorttitle{Galaxy morphology catalog}
\shortauthors{Kuminski \& Shamir}

\begin{document}



\title{A catalog of automatically detected ring galaxy candidates in PanSTARRS}


\author{Ian Timmis 
and Lior Shamir$^{*}$ 
}
\affil{Lawrence Technological University, MI 48075  \\ $^{*}$email: lshamir@mtu.edu
}




\begin{abstract}
We developed and applied a computer analysis method to detect ring galaxy candidates in the first data release of PanSTARRS. The method works by applying a low-pass filter, followed by dynamic global thresholding to search for closed regions in the binary mask of each galaxy image. Applying the method to $\sim3\cdot10^6$ PanSTARRS galaxy images produced a catalog of 185 ring galaxy candidates based on their visual appearance.

\end{abstract}





\section{Introduction}
\label{introduction}


Ring galaxies are rare irregular galaxies that are not on the Hubble classifications scheme. \cite{theys1976ring} proposed a separate classification scheme for ring galaxies that includes three sub-classes based on their visual appearance: Empty ring galaxies (RE), Ring galaxies with off-center nucleolus (RN), and ring galaxies with knots or condensations (RK). They also identified that most, although not all, ring galaxies have a companion \citep{theys1977ring}.  \cite{few1986ring} separated ring galaxies into two sub-classes: `'P-type'' rings, which have a knotty structure or an off-center nucleolus, and ``O-rings'', characterized by a smooth ring structure and a centered nucleolus. 

Ring galaxies can be identified as polar rings \citep{whitmore1990new,maccio2005origin,reshetnikov1997global,finkelman2012polar,reshetnikov2015polar}, collisional rings \citep{appleton1996collisional}, bar-driven or tidially-driven resonance rings \citep{buta2000resonance}, ringed barred spiral galaxies \citep{buta2001dynamics} and Hoag-type objects \citep{longo2012morphological}. The ``Hoag's Object'' \citep{hoag1950peculiar,brosch85,schweizer1987structure} was discovered in 1950, and its discovery was followed by the identification of other ring galaxies. 

Catalogs of ring galaxies were created in the past by manual observation. The early \cite{arp1966atlas} catalog of peculiar galaxies contains two galaxies with visual appearance of an empty ring. The catalog of southern peculiar galaxies \citep{arp1988catalogue} includes 69 systems identified as rings.  \cite{whitmore1990new} compiled a list of 157 polar ring galaxy candidates, and about half a dozen of these objects were confirmed as polar ring galaxies by kinematic follow-up observations \citep{finkelman2012polar}. \cite{madore2009atlas} prepared an atlas of collisional ring galaxies. \cite{garcia2015initiating} discovered 16 polar ring galaxy candidates. \cite{buta1995catalog} created a catalog of Southern ring galaxies. \cite{moiseev2011new} used crowdsourcing and non-scientists volunteers to prepare a catalog of ring galaxy candidates through the Galaxy Zoo citizen science campaign.

While manual analysis performed by expert or citizen scientists has provided useful catalogs of ring galaxies, the rapidly increasing data acquisition power of digital sky surveys such as the Large Synoptic Survey Telescopes (LSST) can potentially allow the identification of a very large number of ring galaxies among a total of billions of astronomical objects. Due to the large size of these databases, effective identification of these objects requires automation, leading to the development of automatic methods of identifying peculiar objects in large databases of galaxy images \citep{shamir2012automatic,shamir2014automatic,shamir2016morphology}. Here we describe an automatic image analysis method that can identify ring galaxies, and apply the method to mine through $\sim3\cdot10^6$ galaxies imaged by the Panoramic Survey Telescope and Rapid Response System  \citep{hodapp2004design,flewelling2016pan,chambers2016pan} to compile a catalog of ring galaxy candidates.


\section{Methods}
\label{method}

\subsection{Data}
\label{data}

The dataset was obtained from the Panoramic Survey Telescope and Rapid Response System (PanSTARRS) first data release \citep{hodapp2004design,flewelling2016pan,chambers2016pan}. The initial dataset includes 3,053,831 objects with r magnitude of less than 19. To avoid stars, the dataset included 2,394,452 objects identified as extended sources in all bands, and 659,379 additional objects that were not identified as extended objects in all bands, but their PSF i magnitude subtracted by their Kron i magnitude was larger than 0.05, and their r Petrosian radius was larger than 5.5''. Objects that were identified as artifacts, has a brighter neighbor, defect, double PSF, or a blend in any of the bands were excluded from the dataset, as such objects require time to download and process while significantly increasing the false positive rate.

The images were then downloaded via the PanSTARRS {\it cutout} service as 120$\times$120 JPG images, in a process similar to the image download done in \citep{kum16}. The JPG images that were downloaded and analyzed were the g band images, as the y/i/g color images were in many cases noisy, and did not allow effective automatic analysis. To avoid pressure on the PanSTARRS web server, one image was downloaded at a time, and therefore the processes required 62 days to complete. 

The initial scale was set to 0.25'' per pixel. As done in \citep{kum16}, after each image was downloaded, all pixels located on the edge of the frame with grayscale value higher than 125 were counted. If the number of these pixels was 25\% or more of the total number of pixels on the edge the scale was increased by 0.05'', and the image was downloaded again. That was repeated until the number of foreground pixels on the edge was lower than 25\% of the total edge pixels. The change of the scale assisted in analyzing objects that are initially too large to fit in a 120$\times$120 image of the initial 0.25'' per pixel scale.

Images that contain substantial noise or artifacts are difficult to analyze correctly, and can trigger false positives as will be explained in Section~\ref{image_analysis}. Due to the large scale of the initial dataset, even a low rate of false detections can lead to an unmanageable resulting dataset. Because compression algorithms are more efficient when the signal is smooth, clean images of real galaxies tend to have a smaller compressed file size, and therefore artifacts and noisy images can be rejected by their compressed file size \citep{kum16}. 
Table~\ref{artifacts} shows examples of galaxy images and their file sizes. Based on empirical observations, a threshold was set so that only images with file sizes of less than 5.5KB were analyzed, and larger files were rejected.

\begin{table}[h!]
\begin{center}
\caption{Examples of clean galaxy images and artifacts or noisy images in PanSTARRS. The file size provides a simple mechanism to reject noisy images.}
\label{artifacts}
\begin{tabular}{lll}
\tableline\tableline
PanSTARRS  & File size & Image \\
object ID      &  (KB)     & \\
\hline
102230806134866752 & 9.40 & \includegraphics[angle=00,scale=.50]{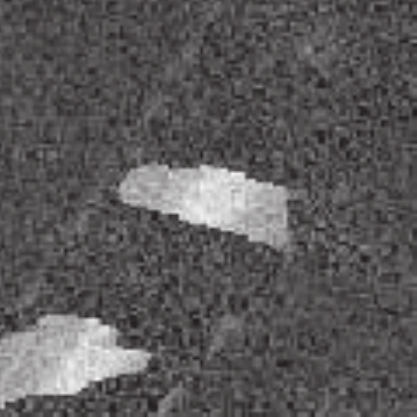} \\
103480451533225122 & 9.58 & \includegraphics[angle=00,scale=.50]{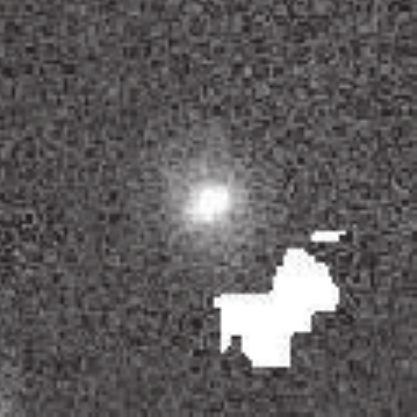} \\
103570759842751155 & 9.43 & \includegraphics[angle=00,scale=.50]{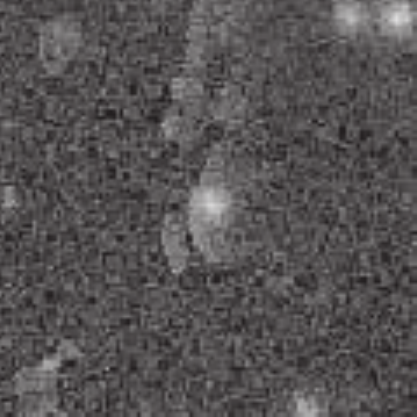} \\
100840464055080903 & 3.17 & \includegraphics[angle=00,scale=.50]{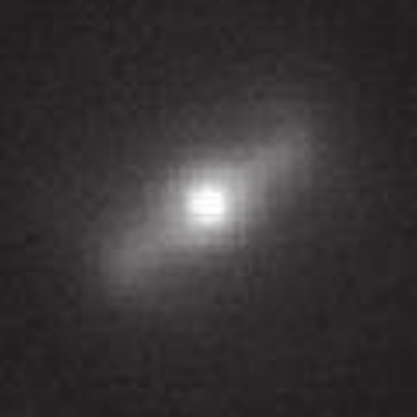} \\
104720155726185389 & 3.10 & \includegraphics[angle=00,scale=.50]{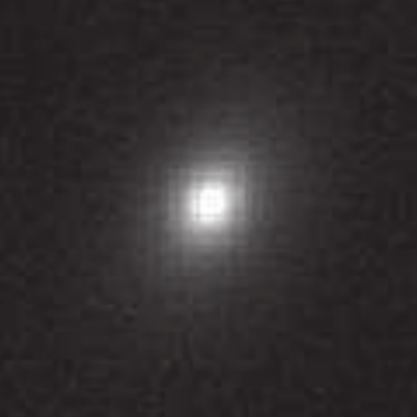} \\
104941422843081464 & 3.88 & \includegraphics[angle=00,scale=.50]{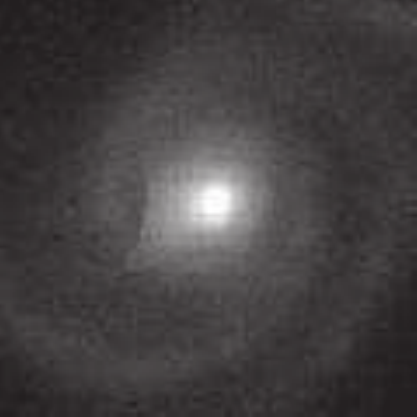} \\
\tableline
\tableline
\end{tabular}
\end{center}
\end{table}

\subsection{Galaxy image analysis}
\label{image_analysis}

\begin {figure} [h]
\centering
\includegraphics[width=0.5\textwidth]{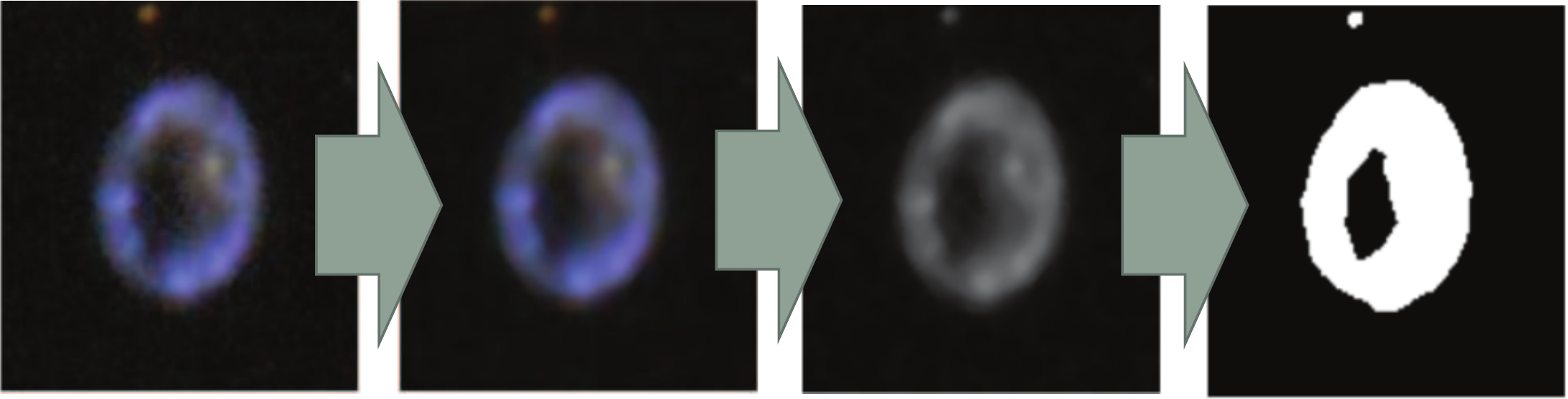}
\caption{The stages in converting the original image into its binary map.}
\label{simplification}
\end {figure}

Each image is smoothed by utilizing a median filter with window size of 5$\times$5 to facilitate noise reduction, and converted to grayscale. The image is then converted into its binary mask using a dynamic threshold. The dynamic threshold starts with a minimum of 30, and is incremented iteratively until it reaches the gray level of 200. The conversion of the original ring galaxy into a binary map is displayed by Figure~\ref{simplification}.

\begin {figure}[h]
\centering
\includegraphics[width=0.5\textwidth]{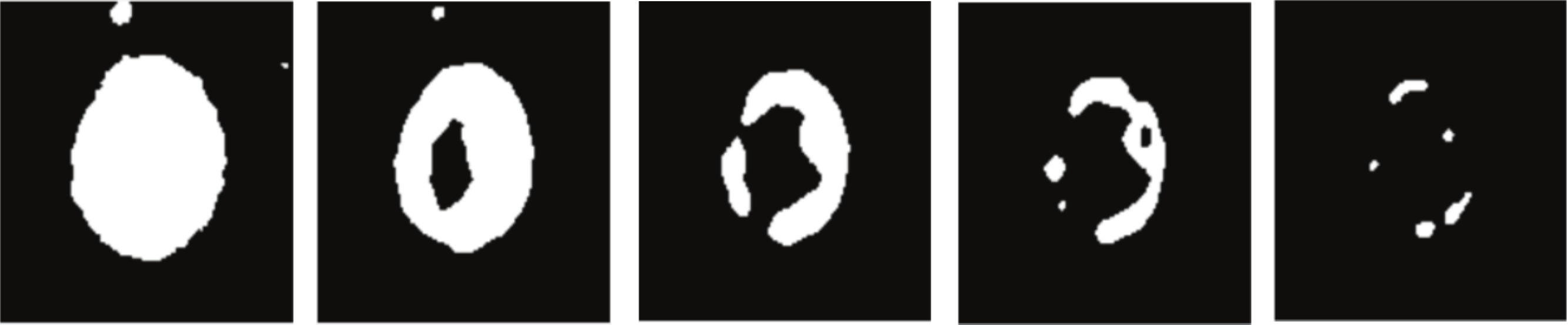}
\caption{Dynamic thresholding binary maps. The Images show binary maps with thresholds of 20 (left), 44, 83, 99, and 115. The figure shows that when using a graylevel of 44 the ring is identified in the binary mask, while other graylevel thresholds show no ring.}
\label{thresholding}
\end {figure}

For each threshold level the binary mask is computed, and a search for a ring inside the foreground is done using a Flood Fill algorithm \citep{asundi1998fast}. Flood fill is an algorithm typically associated with the ``bucket fill'' tool in painting programs. Here we used a stack-based 4-connected version of the flood fill algorithm, which is a non-recursive process starting with an initial pixel and then analyzes the four pixels surrounding it. Each of these four pixels is flagged, and then the neighbors of each of them are also added. That continues until all pixels are flagged, or no neighbors with value of 0 remain. In that case it is determined that no path of pixels of value 0 to the edge exist, and therefore the image is suspected as a ring galaxy. However, if a pixel that is on the edge of the image is flagged, the algorithm stops and it is determined that no ring exists in that graylevel threshold.

\begin {figure}[h]
\centering
\includegraphics[width=0.5\textwidth]{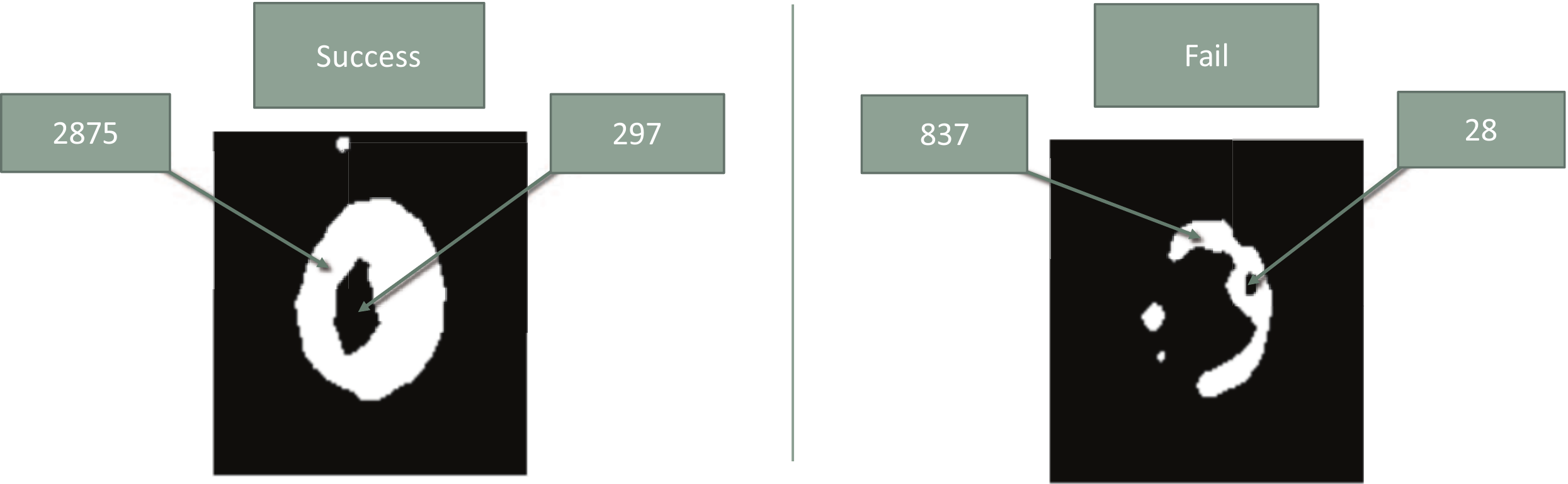}
\caption{Comparison of the size of the closed area to the size of the foreground.}
\label{negative_positive}
\end {figure}

The flood fill algorithm is applied for each pixel in the binary mask. If the flood fill algorithm finishes without reaching a pixel that is on the edge of the frame, the number of pixels in the closed area are counted, and divided by the number of foreground pixels. If the number of pixels in the closed area is less than 10\% of the number of foreground pixels, it is assumed that the closed area is too small to be considered a ring galaxy. Figure \ref{negative_positive} shows an example of closed areas in the binary mask that can be considered candidate rings (left), and small areas in the binary mask of the same image that are merely local grayscale variations (right).

Processing of a small 120$\times$120 galaxy image using a single core of an Intel Xeon E5-1650 requires $\sim$2.1 seconds to complete.




\subsection{False detections}
\label{false_detections}

\begin {figure}[h]
\centering
\includegraphics[scale=0.55]{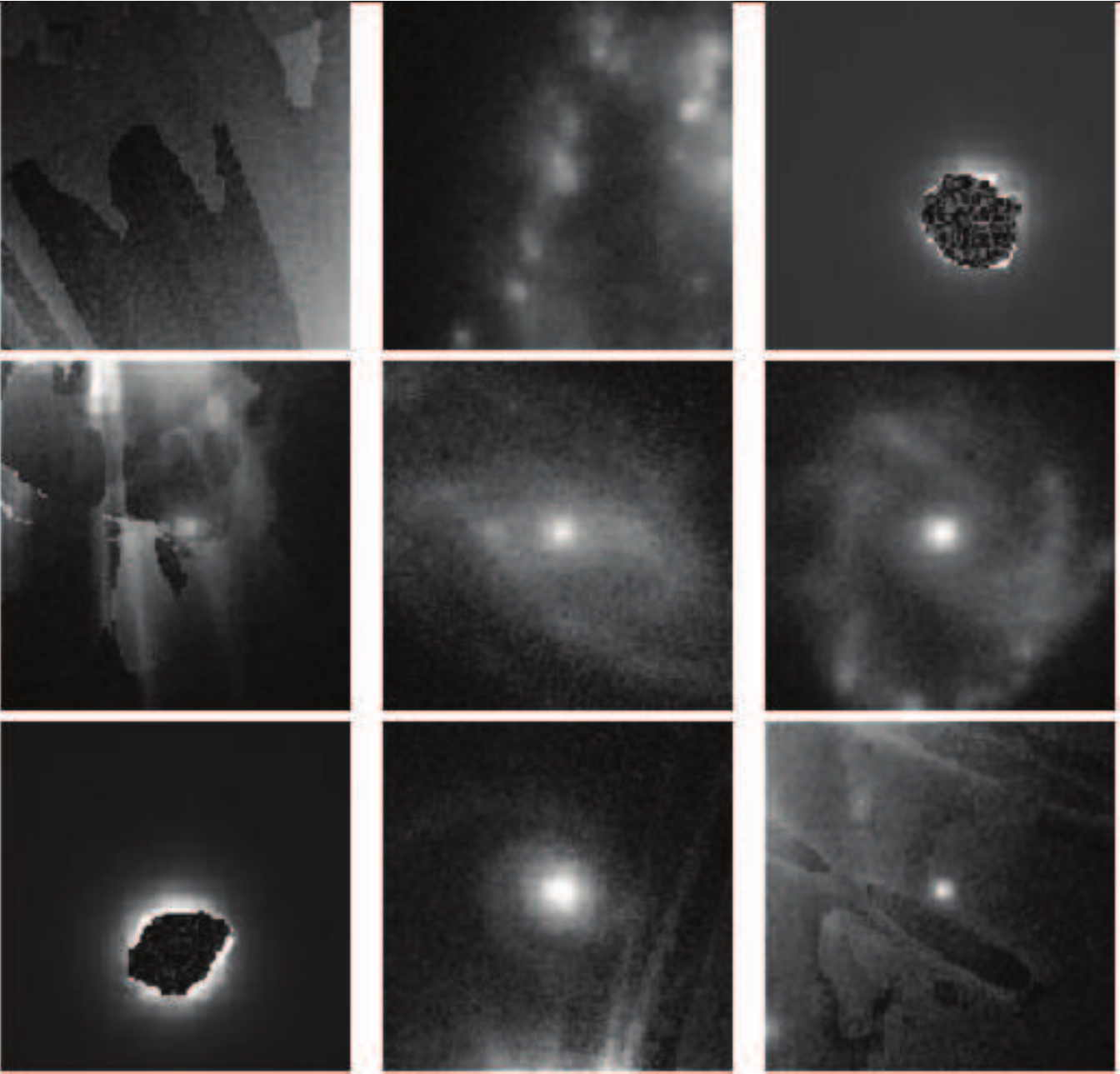}
\caption{Examples of false detections of galaxies as rings}
\label{false_detection}
\end {figure}

When mining through a very large number of galaxies, even a small rate of false detections can lead to an unmanageable database. Of over three million images that were tested, the algorithm detected 2490 galaxies in which manual inspection showed no ring. These galaxies included artifacts, saturated objects, and regular galaxies. Figure~\ref{false_detection} shows examples of false detections of the algorithm. As can be seen in the example images, overlapping arms or stars nearby a spiral galaxy can lead to false detections. Saturated objects can also be mistakenly identified as rings. However, these objects are fairly rare, and the false detection rate is less than 0.1\% of the initial set of galaxies.

Another aspect related to false detection is confusion between ring galaxies and ringed disk galaxies \citep{buta2013galaxy}. That difference, however, is more difficult to identify automatically, as ringed disk galaxies feature ring-like structures, normally as nuclear, inner, or outer rings.

\section{Ring galaxy candidates}
\label{catalog}

The ring galaxy candidates that were detected with their right ascension and declination coordinates are shown in Table~\ref{ring_galaxies}, ordered by the right ascension. 


\onecolumn


\begin{longtable}[ht]{|l|c|c|c|}



  \hline
   No. &  Object ID & RA ($^o$) & Dec ($^o$)  \\ 
  \hline

1 &  102530019896912092 &  1.9897 &  -4.5569 \\ 

2  &  98880036039558395 & 3.6039 & -7.5933 \\
3 &  135050044989180240 &  4.4989 &  22.5414 \\ 
4 &  138120051156912877 &  5.1157 &  25.1019 \\ 
5 &  72870063744005992 &  6.3744 &  -29.2703 \\ 

6 &  120560064695882350 &  6.4696 &  10.4682 \\ 
7  & 100400085662218636 & 8.5661 & -6.3265 \\
8 &  102310089038629907 &  8.9039 &  -4.7336 \\ 
9 &  103950089186978353 &  8.9186 &  -3.3684 \\ 

10 &  137290090541388746 &  9.0541 &  24.4152 \\ 
11 &  115220097713265663 &  9.7715 &  6.0210 \\ 
12 &  109660114619281831 &  11.4619 &  1.3845 \\ 
13 &  111110125282677623 &  12.5283 &  2.5977 \\ 

14 &  120890127618103802 &  12.7618 &  10.7444 \\ 
15  &  92360151168438698 & 15.1168 & -13.0264 \\  
16 &  119290184439583690 &  18.4440 &  9.4110 \\ 
17 &  120590187339699694 &  18.7340 &  10.4993 \\ 
18 &  105620208384294429 &  20.8384 &  -1.9799 \\ 
19 & 112540211443478684 &  21.1444 &  3.7900 \\ 
20 &  108010239026888389 &  23.9027 &  0.0150 \\ 

21 &  130250272062931167 &  27.2063 &  18.5422 \\ 
22 &  152300273312165338 &  27.3313 &  36.9206 \\ 
23 &  136880303174084441 &  30.3172 &  24.0699 \\ 
24 &  117300373539209184 &  37.3539 &  7.7573 \\ 

25 & 144270387141449300 &  38.7141 &  30.2323 \\ 
26 &  98420402083859546 &  40.2084 &  -7.9757 \\ 
27 &  148070464186033529 &  46.4186 &  33.3941 \\ 

28  & 102120545319045128 & 54.5319 & -4.8960 \\
29 &  89730546881066762 &  54.6881 &  -15.2197 \\ 
30 & 105770557953197395 &  55.7953 &  -1.8525 \\ 

31 &  94400582354418830 &  58.2354 &  -11.3263 \\ 

32 &  98300585149137156 &  58.5149 &  -8.0777 \\ 

33 &  103680672704100823 &  67.2704 &  -3.5997 \\ 
34 &  87960731606800107 &  73.1606 &  -16.7 \\ 
35 &  110590740360460369 &  74.0359 &  2.1585 \\ 
36 &  93170740765876164 &  74.0766 &  -12.3534 \\
37 &  90050743863589139 &  74.3863 & -14.951 \\ 
38  &  90040746028548781 & 74.6028 & -14.9596 \\    
39 &  112280754250309794 &  75.4252 &  3.5745 \\ 
40 &  108480776785245228 &  77.6784 &  0.4040 \\ 

41 &  98820784238446846 &  78.4238 &  -7.6447 \\ 
42 &  85450837400860065 &  83.74 &  -18.7919 \\ 

 43 &  77350864221236846 &  86.4221 &  -25.5362 \\ 
 44 &  179561088508306139 &  108.8508 &  59.6379 \\ 
45 &  154011105647715883 &  110.5648 &  38.3460 \\ 

46 &  194551123628248035 &  112.3629 &  72.1311 \\ 

 47 &  183491156027365911 &  115.6027 &  62.9125 \\ 
 48 &  151781161775920391 &  116.1776 &  36.4832 \\ 
49 &  170881181537928277 &  118.1536 &  52.4061 \\ 
50 &  164611201937762708 &  120.1938 &  47.1767 \\ 
51 & 141001202112940207 &  120.2113 &  27.4997 \\ 
52 &  133621235595071267 &  123.5595 &  21.3506 \\ 

53 &  124791272138223104 &  127.2138 &  13.9938 \\ 

54 &  171231275002484748 &  127.5002 &  52.6951 \\
55 &  109361301796774389 &  130.1797 &  1.1367 \\ 
56 &  190291319360661229 &  131.9360 &  68.5754 \\ 
57 &  188281380666241185 &  138.0666 &  66.9 \\ 
58 &  98371387382707506 &  138.7383 &  -8.0191 \\ 
59 &  125181389138828530 &  138.9139 &  14.3234 \\ 

60 & 130811416883240083 &  141.6883 &  19.0080 \\ 
61 &  171161427964224003 &  142.7964 &  52.6361 \\ 

62 &  82891436370344404 &  143.6370 &  -20.9216 \\ 
63 &  94081470757770101 &  147.0758 &  -11.6001 \\ 
64 &  123371475276486913 &  147.5276 &  12.8137 \\ 
65 &  111101480001954462 &  148.0002 &  2.5867 \\ 

66 & 154241493558096897 &  149.3558 &  38.5386 \\ 
67 & 166351516903793594 &  151.6904 &  48.6275 \\ 
68 & 107721522634947956 &  152.2635 &  -0.2271 \\ 
69 & 141541526163197945 &  152.6163 &  27.9563 \\ 

70 & 151261533456282945 &  153.3457 &  36.0520 \\ 
71 & 181411561600712736 &  156.1601 &  61.1767 \\ 
72 &  160351569982260063 &  156.9982 &  43.6245 \\ 
73 & 123431607654498382 &  160.7655 &  12.8649 \\   
74 &  129391627135806697 &  162.7136 &  17.8302 \\ 
75 &  76451645788803086 &  164.5791 &  -26.2894 \\ 
76 & 98941648591968075 &  164.8592 &  -7.5435 \\ 

77 & 153881653487717633 &  165.3487 &  38.2392 \\ 
78 &  148781654085861195 &  165.4086 &  33.9839 \\ 
79 &  168691656455379447 &  165.6455 &  50.5823 \\ 

80 &  122761662462277308 &  166.2462 &  12.3057 \\ 
81 & 167061663892379757 &  166.3893 &  49.2242 \\ 
82 &  176441677887278299 &  167.7888 &  57.0397 \\ 
83 &  177161677083354182 &  167.7084 &  57.6362 \\ 
84 &  147611692157354849 &  169.2158 &  33.0119 \\ 
85 & 169761694401977726 &  169.4402 &  51.4725 \\ 
86 &  131551702348949298 &  170.2348 &  19.6322 \\ 
87 &  83511707305532699 &  170.7306 &  -20.4065 \\ 
88 & 132441708586612664 &  170.8587 &  20.3684 \\ 
89 &  120061740415357108 &  174.0416 &  10.0555 \\ 
90 &  145631743790845430 &  174.3791 &  31.3623 \\ 

91 & 166121745010278457 &  174.5011 &  48.4398 \\ 
92 &  103021750976199429 &  175.0976 &  -4.1425 \\ 
93 &  140401760507263297 &  176.0507 &  27.0023 \\ 

94 &  182681764076585291 &  176.4077 &  62.2371 \\ 
95 &  171201764688293595 &  176.4689 &  52.6691 \\ 
96 &  103361766457995817 &  176.6458 &  -3.8621 \\ 
97 &  100891767811050892 &  176.7811 &  -5.9246 \\ 
98 &  180361768453241219 &  176.8454 &  60.3004 \\ 
99 &  81531787292064732 &  178.7292 &  -22.0546 \\ 
100 &  146751794344735513 &  179.4344 &  32.2957 \\ 
101 & 126861796962350039 &  179.6962 &  15.7163 \\ 
102 &  120911807953598022 &  180.7954 &  10.7646 \\ 

103 &  125491835740118715 &  183.5740 &  14.5818 \\  
104 & 121741853507871762 &  185.3508 &  11.4511 \\ 
105 & 112761866548340926 &  186.6548 &  3.9671 \\ 
106 &  137771871189921740 &  187.1190 &  24.8093 \\ 

 107 &  114301857795845888 &  185.7796 &  5.2545 \\ 
 108 &  125741873384637652 &  187.3385 &  14.7893 \\ 
 109 &  113901876468607202 &  187.6469 &  4.9223 \\ 
110 & 144871915423938960 &  191.5421 &  30.7321 \\ 

 111 &  131131919673730819 &  191.9674 &  19.2752 \\ 
112 &  152181932753561700 &  193.2755 &  36.8174 \\ 
113 &  102921932899245357 &  193.2899 &  -4.2292 \\ 
114 &  102271941347122534 &  194.1347 &  -4.7732 \\ 
 115 &  164961951082294851 &  195.1082 &  47.4702 \\ 
116 &  119131965074384721 &  196.5074 &  9.2786 \\ 

 117 &  127841971925597729 &  197.1926 &  16.5394 \\ 
118  &  102261975134478601 & 197.5134 & -4.7764 \\
119 &  107441989471515672 &  198.9471 &  -0.4623 \\ 
120 &  105281992642316849 &  199.2642 &  -2.2613 \\ 
121 &  95721996172550718 &  199.6172 &  -10.2331 \\ 
122 &  133081997942710534 &  199.7943 &  20.9000 \\ 

123 &  181722026514721806 &  202.6515 &  61.4342 \\ 
124 &  174802054136252600 &  205.4136 &  55.6682 \\ 

125 &  125292060017968309 &  206.0018 &  14.4148 \\ 
126 &  103632081566171282 &  208.1566 &  -3.6409 \\ 
127 &  127392083085689351 &  208.3086 &  16.1657 \\ 

128 &  77702093105650034 &  209.3106 &  -25.2503 \\ 

129 &  184722100380214976 &  210.0380 &  63.9369 \\ 
130 &  112512111057222057 &  211.1057 &  3.7597 \\ 
131 &  123332115027789738 &  211.5028 &  12.7828 \\ 
132 &  123362118086413970 &  211.8086 &  12.8029 \\ 
133 &  103052137786520187 &  213.7787 &  -4.1252 \\ 

134 &  113472161524401485 &  216.1525 &  4.5592 \\ 
135 &  109492177573061925 &  217.7573 &  1.2429 \\ 

136 &  84542211439173056 & 221.1439 & -19.5477 \\
137 &  116152228525787253 &  222.8526 &  6.7973 \\ 

138 &  183352242303877381 &  224.2304 &  62.7972 \\ 

139 &  131032259389706416 &  225.9390 &  19.1966 \\ 

140 &  148062302119596631 &  230.2120 &  33.3884 \\ 
141 &  183392321197746552 &  232.1198 &  62.8298 \\ 
142 &  148432328285931174 &  232.8286 &  33.6921 \\ 
 
 143 &  158172331153014314 &  233.1150 &  41.8113 \\ 
 144 &  148312333416204412 &  233.3416 &  33.5948 \\ 
145 &  143522339208520736 &  233.9209 &  29.6001 \\ 
 
146 &  129292352520407811 &  235.2520 &  17.7477 \\ 
147 &  129462367458987309 &  236.7460 &  17.8890 \\ 

148 &  202792388527621833 &  238.8528 &  78.9925 \\ 

149 &  132902394059167588 &  239.4059 &  20.7559 \\
150 &  131452396300467354 &  239.6301 &  19.5474 \\  
151 &  146322413429998230 &  241.3430 &  31.9397 \\ 
152 &  110832424057543879 &  242.4057 &  2.3612 \\ 
153 &  138262424254535667 &  242.4254 &  25.2209 \\ 

154 &  108572432090906806 &  243.2091 &  0.4803 \\ 
155 &  150852441278310269 &  244.1278 &  35.7081 \\ 
156 &  144432453827808671 &  245.3828 &  30.3651 \\        
 
157 &  120402497965084423 &  249.7965 &  10.3366 \\ 
 
158 &  164882506974275102 &  250.6974 &  47.4037 \\ 
159 &  182962541826568261 &  254.1827 &  62.4729 \\ 
160 &  137392563849799521 &  256.3850 &  24.4991 \\ 
161 &  142632581066552766 &  258.1067 &  28.8602 \\ 
162 &  178582585376172798 &  258.5377 &  58.8184 \\ 
163 &  154302667799896469 &  266.7800 &  38.5882 \\ 
164 &  162442713696719955 &  271.3698 &  45.3742 \\ 
165 &  187092713680201538 &  271.3680 &  65.9089 \\ 

166 &  180632732359767042 &  273.2359 &  60.5303 \\ 

167 &  85913110560453012 &  311.0560 &  -18.4061 \\ 
168 &  90603151049933857 &  315.1050 &  -14.4971 \\ 
169 &  113963209792571680 &  320.9793 &  4.9677 \\ 
170 &  102753218509133054 &  321.8509 &  -4.3728 \\ 

171 &  121663222491624631 &  322.2492 &  11.3868 \\ 

172 &  87453246504240833 &  324.6504 &  -17.1246 \\ 

173  &  82723283048699480 & 328.3048 & -21.0590 \\
174 &  136183298529911556 &  329.8530 &  23.4842 \\ 
175 &  82903312563109880 &  331.2564 &  -20.9087 \\ 
176 &  131603342959906588 &  334.2960 &  19.6717 \\ 

177 &  78253370037949323 &  337.0038 &  -24.7842 \\ 

178 &  127173454713128088 &  345.4713 &  15.9813 \\ 
179 &  124773504516095667 &  350.4516 &  13.9793 \\ 

180 &  147543515744331454 &  351.5745 &  32.9506 \\ 
181 &  102293536494361637 &  353.6494 &  -4.7573 \\ 

182 &  132133578627738343 &  357.8628 &  20.1149 \\ 

183 &  118623589791533531 &  358.9792 &  8.8525 \\ 
184 &  136013594740073572 &  359.4740 &  23.3442 \\ 
185 &  120513599153160081 &  359.9153 &  10.4247 \\

\hline
\caption{Ring galaxy candidates identified automatically} \\
\label{ring_galaxies}
\end{longtable}

\twocolumn

\begin {figure*}[h]
\centering
\includegraphics[scale=0.38]{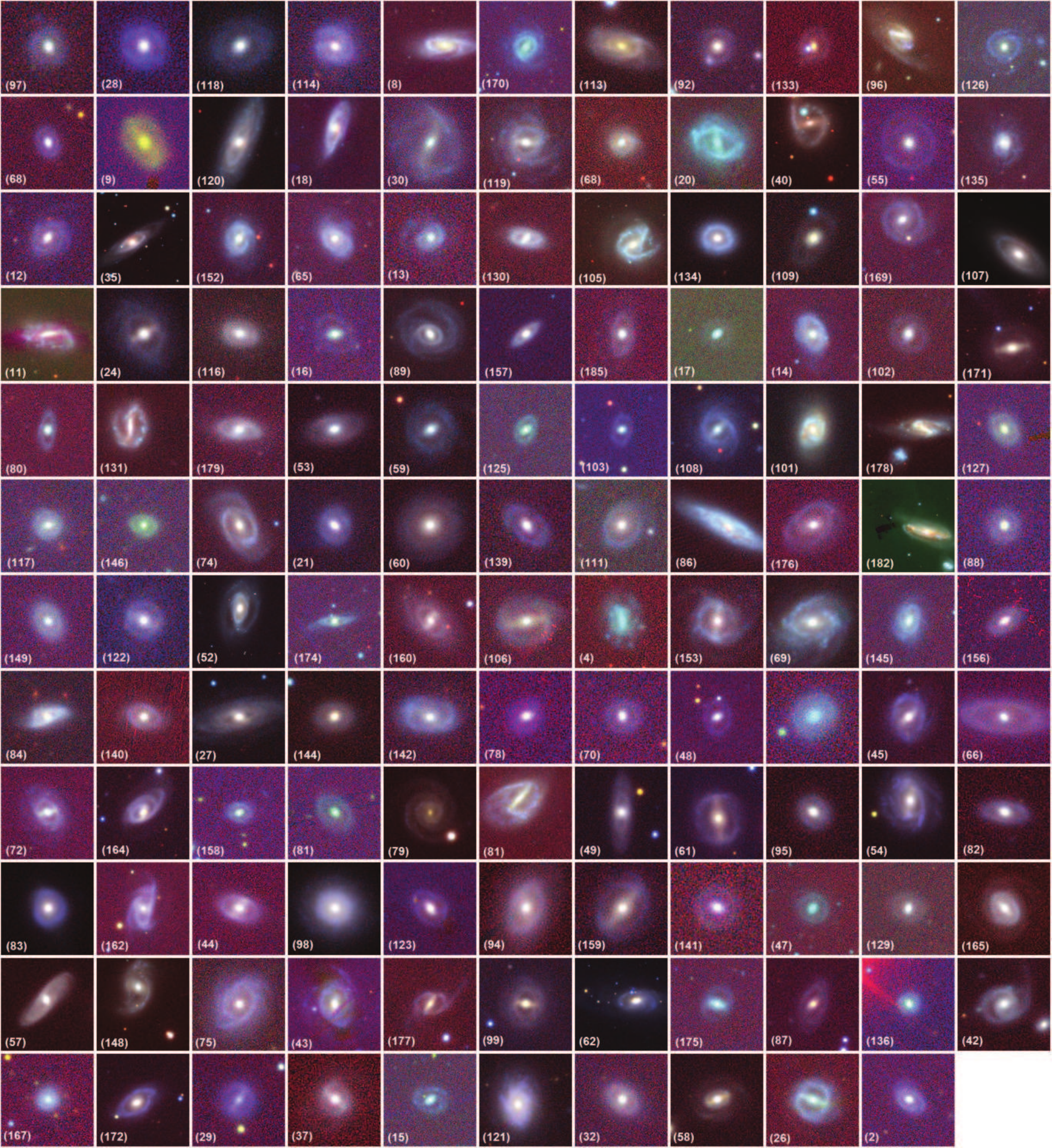}
\caption{PanSTARRS images of resonance ring candidates in ordinary galaxies.}
\label{resonance}
\end {figure*}

\begin {figure*}[h]
\centering
\includegraphics[scale=0.38]{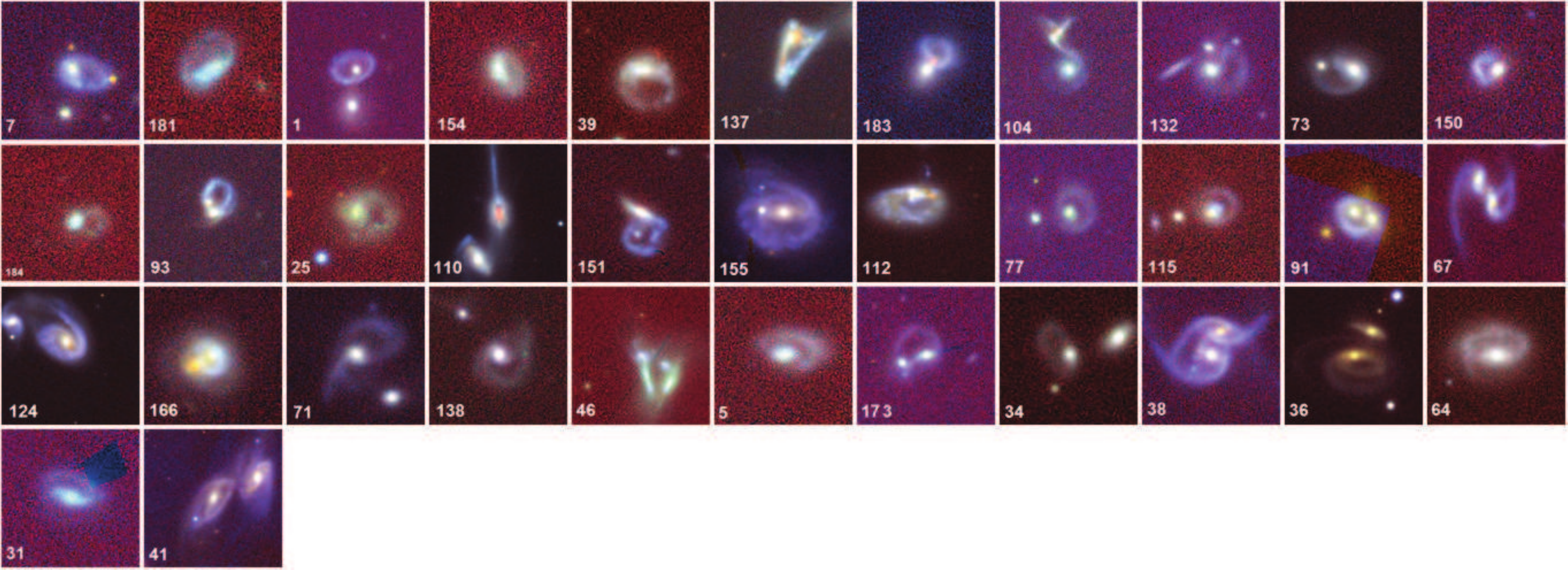}
\caption{PanSTARRS images of collisional ring candidates.}
\label{collisional}
\end {figure*}

\begin {figure*}[h]
\centering
\includegraphics[scale=0.38]{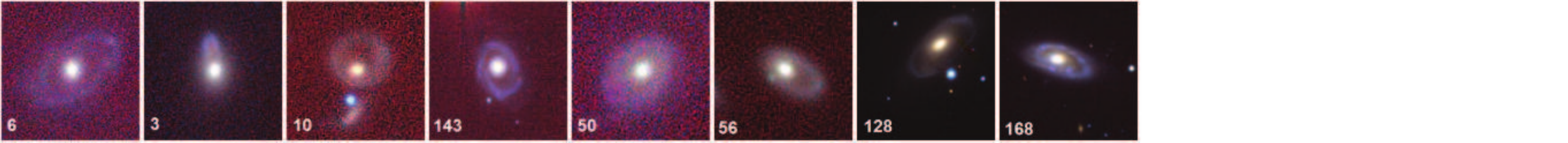}
\caption{PanSTARRS images of ring galaxies with off-center nucleus.}
\label{off_center}
\end {figure*}

\begin {figure*}[h]
\centering
\includegraphics[scale=0.38]{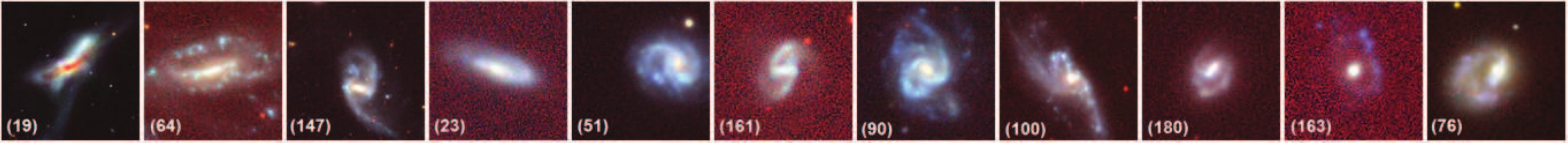}
\caption{PanSTARRS images of other galaxies not included in Figures~\ref{resonance},~\ref{collisional}, and~\ref{off_center}.}
\label{others}
\end {figure*}

PanSTARRS images of the galaxies are displayed by Figures~\ref{resonance},~\ref{collisional},~\ref{off_center}, and~\ref{others}.

\subsection{Comparison to Madore collisional ring Atlas}
\label{madore}

To assess the completeness of the catalog, the detected galaxies were compared to the \cite{madore2009atlas} catalog of collisional ring galaxies. The  objects listed in Table~\ref{madore} are objects from that catalog that are inside the footprint of PanSTARRS Data Release 1. The table shows the corresponding PanSTARRS object ID, and the Kron magnitude measured on the r band, which is used as a criteria for inclusion of objects in the initial dataset as described in Section~\ref{method}. When multiple objects IDs are associated with the same extended source, the selected object ID is the object that its photometric information is the closest to the photometry threshold for selecting the objects as described in Section~\ref{data}. The PSF i magnitude subtracted by the Kron i magnitude was also used as a method of filtering objects that are not galaxies, as well as the identification of the object as an extended sources in the g,r,i and z bands. When iPSFMag-iKronMag is larger than 1000 it means that one of the iKronMag reading was bad, leading to a -999 value. The table also shows what objects were detected as ring by the algorithm by downloading the image directly from PanSTARRS server and running the algorithm.

\begin{table*}[h!]
{\scriptsize
\begin{center}
\caption{Collisional ring galaxies from \citep{madore2009atlas} that are inside the footprint of PanSTARRS DR1.}
\label{madore}
\begin{tabular}{lllllllll}
\tableline\tableline
Catalog & PanSTARRS  & RA  & DEC &  Included                         &  r Kron        & iKronMag-          &  extended               & Ring detected  \\
name     & object ID      &      &         &  in Table~\ref{ring_galaxies} &  magnitude &  iPSFMag            &  source bands   & by the algorithm \\
\hline
Arp 146 & 100030016843707022 & 1.6841 & -6.635    & No  &  15.97    & 3.34    & r,i,z & No \\  
Arp 318 & 95800323797258621 & 32.379 & -10.159    & No  & 22.492 & 1022.611 & -- & n.a. \\
Arp 10   & 114770346040298405 & 34.609   & 5.653    & No  & -999     &   1020.9     & -- & Yes \\
Arp 273 & 155240353777382539  & 35.377 & 39.366   & No &   15.42  & 1021.098  &  g, r & No  \\ 
Arp 145 & 157650357813012712 & 35.785 & 41.372    & No &  16.59   & 4.68          & g,r,i,z &    No \\
NGC 985 & 97450386522587254 & 38.657 & -8.787     & No &  15.7 & 4.12 & g,r,i,z & No \\ 
Arp 118 & 107770437986659641 & 43.795 & -0.181    & No  & 13.2 & 5.22 & r,i,z & Yes \\ 
Arp 147 & 109570478265116474 & 47.829 & 1.315     & No  & 16.89  & 0.86 & r,i,z & No \\
Arp 219 & 105450549711618738 & 54.975 & -2.118    & No & 17.14 & 3.8 & r,i,z & Yes \\
Arp 141 &  196161085846518933 & 108.585 & 73.477 & No  & -999 & 1020.056 & g & No \\
Arp 143 & 144021167270940513 & 116.723 & 30.019 & No & 20.18 & 0.98 &  g,r,i,z & No \\
NGC 2793 & 149311391915893056 & 139.197 & 34.430 & No & -999    & 0     & z & No  \\
Arp 107    & 144071630704802930 & 163.069 & 30.065 & No & 16.87 & 3.8  & r,i &  No \\ 
Arp 148    & 157011659698568092 & 165.972 & 40.849 & No & 17.14 & 0 & g,r & Yes \\
VII Zw 466 & 187681880151285551 & 188.018 & 66.404 & No   & 16.76 & 2.73   &  g,r,i & Yes \\    
NGC 4774   & 152191932750252225 & 193.275 & 36.823 & Yes &  15.459 & 1.7 & g,r,i,z & Yes \\ 
Arp 150      & 119403498802526691 & 349.880 & 9.505 & No & 15.18 & 4.85 & g,r,i,z & No \\
\tableline
\tableline
\end{tabular}
\end{center}
}
\end{table*}

With the exception of NGC 4774, the objects are not included in Table~\ref{ring_galaxies}. The reason for the exclusion of these objects from the catalog can be the inability of the algorithm to detect them, as in the case of Arp 145, Arp 146, NGC 985, and Arp 150. Arp 145 is a relatively dim ring, and in the case of Arp150 and Arp 146 the ring is not full, and therefore the method failed to detect it due to an opening in the ring that allows the flood fill ``escape'' from the ring and reach the edge of the frame. These systems were included in the initial list of galaxies, but were not detected due to the inability of the method to detect them.

\begin {figure*}[h]
\centering
\includegraphics[scale=1.0]{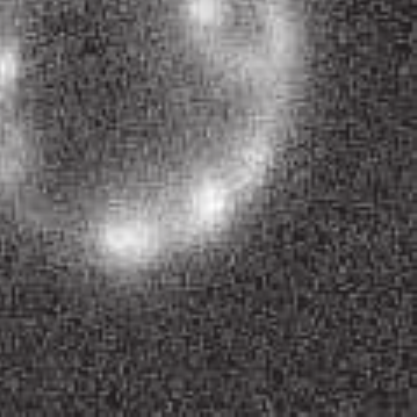}
\caption{The 120$\times$120 image of VII Zw 466 as downloaded for the initial list of galaxies described in Section~\ref{data}}
\label{zw466}
\end {figure*}

An interesting case is the object VII Zw 466, which was used as the object for demonstrating the algorithm in Section~\ref{image_analysis}, but was not detected when applying the algorithm to the PanSTARRS images. The reason is that the PanSTARRS photometric object associated with it that was in the initial list is not the center of the ring, as shown by Figure~\ref{zw466}. Because the object was not centered, the full ring was outside of the frame, and due to the few bright pixels on the frame the object was not identified as too large, and was therefore not re-scaled as described in Section~\ref{data}.

The other objects were not included in the initial list of objects described in Section~\ref{data}. For instance, Arp 318 is a group of ``faint, diffuse streamers, peculiar galaxies'' \citep{arp1966atlas}, and as such is outside the scope of objects that can be identified by the algorithm described in Section~\ref{method}. It also has bad rKronMag measurement and was not detected as an extended source in any of the bands. Arp 10 also has bad rKronMag measurement, and Arp 273 has bad iKronMag measurement and was identified as an extended source only in two bands, and therefore did not meet the criteria for the initial list of galaxies described in Section~\ref{data}. When testing the images, six out of the 17 systems were identified as rings by the algorithm. That shows that even if all PanSTARRS galaxies were tested, many more relevant system would still be hidden in the PanSTARRS database.

\section{Conclusion}
\label{conclusion}

Autonomous sky surveys have enabled the acquisition of very large databases of image and other data, substantially increasing the discovery power of ground and space-based telescopes. To utilize the discovery power and turn these data into scientific discoveries, it is required to apply computational methods that can mine these very large databases. Since a substantial part of these data are in the form of images, full analysis of the data requires image analysis methodology. Here we use a simple and fast automatic image analysis method and apply it to the PanSTARRS first data release to detect ring galaxy candidates. Despite the simple nature of the image analysis method, it can find ring galaxies that are highly difficult to find without using automation, and it is sufficiently fast to be applied to much larger databases such as LSST.

That shows that it is reasonable to assume that many more objects with ring structure could exist in PanSTARRS DR1, and were not detected in this experiment. Identifying all objects in PanSTARRS will require the improvement of the algorithm so that it can better handle ``edge'' cases, but also analysis of a larger dataset of PanSTARRS objects, as it is possible that many relevant objects did not meet the criteria for the initial data reduction. The PanSTARRS photometric pipeline can in some cases provide bad measurement values (e.g., ``-999'') or fail to identify an extended source, leaving the object outside of the initial list of galaxies. The initial data reduction is required for reducing the very large PanSTARRS dataset of over $3\times10^9$ objects to a ``manageable'' number of galaxies that can be downloaded and analyzed. Some objects such as Arp 138 are too large or have morphology that cannot be identified by the proposed algorithm. However, this study shows that applying a first step of automatic image analysis can identify objects that would require substantial labor to identify manually.

\section{Acknowledgments}

We would like to thank the anonymous  reviewer for the insightful comments that helped to improve the manuscript. This study was supported by NSF grant IIS-1546079.

The Pan-STARRS1 Surveys (PS1) and the PS1 public science archive have been made possible through contributions by the Institute for Astronomy, the University of Hawaii, the Pan-STARRS Project Office, the Max-Planck Society and its participating institutes, the Max Planck Institute for Astronomy, Heidelberg and the Max Planck Institute for Extraterrestrial Physics, Garching, The Johns Hopkins University, Durham University, the University of Edinburgh, the Queen's University Belfast, the Harvard-Smithsonian Center for Astrophysics, the Las Cumbres Observatory Global Telescope Network Incorporated, the National Central University of Taiwan, the Space Telescope Science Institute, the National Aeronautics and Space Administration under Grant No. NNX08AR22G issued through the Planetary Science Division of the NASA Science Mission Directorate, the National Science Foundation Grant No. AST-1238877, the University of Maryland, Eotvos Lorand University (ELTE), the Los Alamos National Laboratory, and the Gordon and Betty Moore Foundation.




\bibliographystyle{apalike} 
\bibliography{ms}

\clearpage



\end{document}